\documentclass{WileyMSP-template}
\usepackage[utf8]{inputenc} 
\usepackage[T1]{fontenc}    
\usepackage[british]{babel} 

\usepackage[sfdefault,condensed]{roboto} 
\usepackage[italic]{mathastext}          

\usepackage[dvipsnames]{xcolor}          
\usepackage{hyperref}                    
\hypersetup{
	colorlinks   = true,                 
	urlcolor     = Orange,               
	linkcolor    = RoyalBlue,            
	citecolor    = RoyalBlue             
}

\usepackage{booktabs}      
\usepackage{multirow}      
\usepackage{adjustbox}     
\usepackage{subfig}        

\usepackage{setspace}      
\usepackage{siunitx}       
\usepackage[version=4]{mhchem}        
\usepackage{miller}        
\usepackage{textgreek} 

\usepackage[numbers,sort&compress]{natbib}

\begin{document}
	
	
	\title{\textbf{Unlocking nanoscale microstructural detail in aluminium alloys through differential phase contrast segmentation in STEM}}
	
	\maketitle
	
	
	\author{Matheus A. Tunes*, Martin Hasenburger, Rostislav Daniel, Oscar M. Prada-Ramirez, Philip Aster, Sebastian Samberger, Thomas M. Kremmer, and Johannes A. Österreicher}
	
	\dedication{Dedicated to Professor Peter J. Uggowitzer (ETH Zürich) on the occasion of his 75th birthday -- whose five decades of devotion to metallurgy and materials science, and boundless passion for collaboration, remain an inspiration for our community.}
	
	\begin{affiliations}
		Prof. Dr. M.A.Tunes, M. Hasenburger, Dr. P. Aster, Dr. S. Samberger, and Dr. T.M. Kremmer\\
		Department Metallurgy\\
		Chair of Nonferrous Metallurgy\\
		Laboratory of Metallurgy in Extreme Environments\\
		Montanuniversität Leoben\\ Franz-Josef-Straße 18 \\ 8700 Leoben, Austria \\
		\href{mailto:matheus.tunes@unileoben.ac.at}{matheus.tunes@unileoben.ac.at}\\
		\href{https://x-mat.unileoben.ac.at}{x-mat.unileoben.ac.at}
		
		\
		
		Prof. Dr. R. Daniel \\
		Department Materials Science \\
		Chair Functional Materials and Materials Systems \\
		Montanuniversität Leoben\\ Franz-Josef-Straße 18 \\ 8700 Leoben, Austria 
		
		\
		
		Dr. O.M. Prada-Ramirez\\
		Universidade de São Paulo \\
		Escola Politécnica \\
		Departamento de Engenharia Metalúrgica e de Materiais \\ Av. Prof. Mello Moraes, 2463 \\ 05508-900 São Paulo, Brazil
		
		\
		
		Dr. J.A. Österreicher \\
		LKR Light Metals Technologies \\
		AIT Austrian Institute of Technology \\
		Lamprechtshausenerstrße 61 \\
		5282, Ranshofen, Austria
		
	\end{affiliations}
	

	\begin{abstract}
		\justify
		Differential phase contrast (DPC) imaging in scanning transmission electron microscopy (STEM) maps projected electric fields through the phase sensitivity of segmented low-angle detectors. Although typically applied to atomic-resolution imaging at low beam currents, STEM-DPC is here demonstrated as a rapid micro- and nanoscale image-segmentation tool for materials characterization in advanced aluminium alloys. Decomposition of false-colour DPC micrographs in hue–saturation–value space enables simultaneous identification and quantification of nanoclusters, GP zones, intermediate precipitate phases, dislocation cores, and associated strain fields within a single field of view. The method is demonstrated across multiple alloy systems, including clustering and strain-field mapping in a deformed AlMgZn(Cu) crossover alloy, precipitate identification in a paint-baked automotive sheet alloy, phase-variant segmentation in overaged AA7075-T7, and nanopore and nanoparticle detection in an anodic coating on AA2024-T3. Coupling DPC with neural-network segmentation further enables automated grain-boundary delineation and quantification in nanocrystalline aluminium thin films. Combined with STEM-EDX, DPC-based segmentation enables correlative microstructural analysis, establishing DPC as a rapid complement to techniques such as SPED and 4D-STEM.
	\end{abstract}
    
	\newpage
	\tableofcontents
	
	\section{Introduction}
	\label{sec:intro}
	\justify
	\noindent Aluminium alloys are of historical significance for society due to their high strength-to-weight ratio, corrosion resistance, manufacturability, recyclability, enabling lightweight and durable structures in applications such as aircraft, vehicles, and infrastructure. Over the past century, these alloys -- along with pure aluminium -- have also been at the forefront of major discoveries in metallurgy and materials science, facilitated by the application of transmission electron microscopy (TEM). In 1986, Professor Peter B. Hirsch reviewed the steps taken by his research group towards the first identification of dislocations using diffraction contrast techniques in the TEM -- a milestone achieved in 1956 by analysing pure aluminium foils in a Siemens and Halske Elmiskop TEM operating at 80 kV \cite{hirsch1956lxviii,hirsch1986direct}. Hirsch \textit{et al.} not only confirmed the existence of dislocations by studying pure aluminium, but also documented their motion with \textit{in situ} TEM \cite{hirsch1956lxviii,hirsch1986direct}. Curiously, these experimental discoveries were reported to the scientific community only two decades after Professor Geoffrey I. Taylor had theoretically proposed the existence of dislocations in metals circa 1934 \cite{taylor1934mechanismone}. Although Taylor pointed out the existence of dislocations by analysing plastic-stress-strain relationships in polycrystalline copper \cite{taylor1934mechanismtwo}, Hirsch's group ``directly imaged'' dislocations in aluminium by developing new imaging techniques within emerging TEMs \cite{hirsch1956lxviii,hirsch1986direct}.
	
	Building on the pioneering electron microscopy work on pure aluminium and selected alloys by Hirsch’s team \cite{hirsch1956lxviii,hirsch1986direct} and British scientists \cite{barnes1960nature,mazey1962interstitial,westmacott1962growth,harkness1969critical}, TEM methods have continued to play a pivotal role in guiding the development of materials, and especially new aluminium alloys. With a focus on the research and development of new strengthening strategies through microstructural modification, aluminium alloys were found in 1938 to be strengthened via the precipitation of Guinier-Preston (GP) zones, evidence collected with Laue diffraction -- thus lacking direct visualisation of microstructural features -- from bulk Al-Cu alloys in experiments reported independently by Guinier \cite{guinier1938structure} and Preston \cite{preston1938structure}. GP zones were first observed in the microstructure of an Al-Ag alloy in 1961 by Freise \textit{et al.} using bright-field TEM (BFTEM) \cite{freise1961guinier}. More recently, advancements in high-resolution scanning transmission electron microscopy (HR-STEM) and techniques such as low-angle annular dark-field (LAADF) imaging have allowed Marioara \textit{et al.} \cite{marioara2024atomic} to experimentally reveal the initial stages of GP zone formation, highlighting solute-defect interactions and the aggregation of clusters -- an entity defined as an aggregate of atoms preceding the formation of GP zones and their subsequent evolution to larger (and stabler) precipitates. The experimental work of Marioara and co-workers exploring multiple TEM/STEM methods and signals in the past three decades \cite{andersen1998crystal,marioara2001atomic,marioara2003influence,marioara2005influence,andersen2005crystal,marioara2007effect,vissers2007crystal,torsaeter2010influence,ninive2014detailed,thronsen2021studying,thronsen2022studying,marioara2024atomic} was of paramount importance to consolidate, correct and validate the theoretical understanding \cite{dumitraschkewitz2018clustering} on microstructural design of aluminium alloys via precipitation-hardening, addressing the persistent strength-ductility trade-off in metallurgy \cite{stemper2022potential}. 
	
	Recent advancements in electron microscopy are pushing the boundaries of metallurgy towards the atomic scale. Investigations of a recycled Al-Mg-Si alloy by Bartawi \textit{et al.} have shown via HR-STEM high-angle annular dark-field (HAADF) imaging that the incorporation of impurities such as Cu and Zn from post-consumer scrap is able to alter chemically and morphologically the precipitated phases in recycled Al-Mg-Si alloy variants, which can be highly detrimental to final mechanical and corrosion properties \cite{bartawi2023atomic}. These authors demonstrated that in recycled aluminium alloys, these ``atomic impurities'' can change the crystalline structure of hardening precipitates by creating impurity-driven short-range atomic configurations in the local environment of the precipitate \cite{bartawi2023atomic}. 
    
    Analysis of the work by Bartawi \textit{et al.} suggests that we have reached an era where the application of modern TEM/STEM methods marks the beginning of ``nanometallurgy'' \cite{allioux2020bi,coradini2024unravelling}, where achieving sustainability and circularity in the metallurgical industry -- as recently envisioned by Raabe \textit{et al.}  \cite{raabe2022making} -- strongly relies on examining the microstructure of metallic alloys and their recycled variants at both the micro- and nano-scale, but now especially at the atomic scale, and in this way, modern S/TEMs have become an indispensable tool for metallurgy.
	
	Modern TEM techniques utilise a wide range of conventional and analytical signals generated by electron beam interactions with material samples. Nowadays, detailed post-analysis of these signals plays a critical role in guiding scientists to the discovery of new materials. Two techniques of interest here have attracted the attention of different materials science communities in recent years. One is LAADF imaging in STEM mode, which has shown exceptional effectiveness in detecting nanoscale microstructural features such as nanoclusters and their associated strain fields in advanced aluminium alloys \cite{marioara2024atomic}. These features are often elusive to other S/TEM imaging methods and even some atomic-scale analytical techniques like atom probe tomography (APT), but are critical for understanding the mechanical behaviour of emerging cluster-hardened advanced aluminium alloys \cite{dumitraschkewitz2018clustering,aster2023strain,marioara2024atomic,aster2024unraveling,fang2025unlocking}. 
    
    The second technique is known as differential phase contrast (DPC). When contrast in TEM arises from phase differences between the incident and scattered beams, the resulting imaging method is termed ``phase-contrast'' \cite{carter2016transmission}. DPC operates in STEM mode and uses the LAADF signal and the phase contrast principle to estimate the magnitude and direction of the electric (or magnetic field) at a specific microstructural site within the specimen \cite{hamilton1984differential,amos2003re,pfeiffer2006phase,hornberger2008differential,shibata2012differential}. As phase-contrast imaging in the TEM is mostly applied in the context of high-resolution TEM, the DPC technique has so far been almost exclusively applied to atomic-scale investigations. De Graaf \textit{et al.} \cite{de2020resolving} recently used DPC to directly image stable columns of hydrogen atoms in a metal hydride: a significant milestone harnessed by DPC's high sensitivity to light atomic species \cite{gauquelin2017determining,yucelen2018phase,de2020resolving}. De Graaf \textit{et al.} demonstrated \cite{de2020resolving} that DPC offers superior resolution and clearer imaging compared to the high-resolution annular bright-field method, which was first shown to be effective for detecting hydrogen atoms in yttrium hydride by Ishikawa \textit{et al.}  \cite{ishikawa2011direct}. Parallel to hydrogen detection in the lattice of metal hydrides, DPC has also been extensively applied for the detection of magnetism \cite{chapman1984investigation,chapman1990modified,tunes2021irradiation} and electrostatic potential of atoms \cite{shibata2012differential} in materials at both atomic and nanoscale levels.
	
	In this study, we further develop the DPC technique using a segmented LAADF detector (DF4) to demonstrate its potential for characterising advanced aluminium alloys at both micro and nanoscale resolutions. While previous studies on DPC have predominantly focused on either atomic-scale imaging or magnetic domain detection \cite{hamilton1984differential,chapman1984investigation,chapman1990modified,amos2003re,pfeiffer2006phase,hornberger2008differential,ishikawa2011direct,shibata2012differential,gauquelin2017determining,yucelen2018phase,de2020resolving,tunes2021irradiation}, within the field of electron microscopy, a broader applicability of DPC for functional metal alloys at the nanoscale remains underexplored. This work demonstrates that the DPC technique is equally effective for both micro- and nanoscale investigations, providing a novel approach for advanced materials characterisation in modern TEM. Focusing on recent advancements in aluminium alloys, we demonstrate that DPC allows decomposition of the transmitted electron signal from complex microstructures into independent segments, enabling unprecedentedly precise analysis of nanoclusters, precipitates, defects like dislocation loops, and (qualitatively) their strain fields with high spatial accuracy and resolution. Two additional case studies highlight the technique's innovative aspects and potential to investigate thermodynamic equilibrium and corrosion mechanisms in aerospace-grade aluminium alloys. We also demonstrate that, when coupled with computational neural network methods for microstructural post-analysis, STEM-DPC imaging allows accurate and statistically significant grain size estimation in a way similar to more complex techniques. Finally, we discuss the integration of DPC with 4D-STEM techniques for future research directions.

    \section{Differential phase-contrast imaging}
	\label{sec:DPCprinciple}
	
	\noindent As noted by Williams and Carter \cite{carter2016transmission}, although phase-contrast imaging is often regarded as \textit{``synonymous with high-resolution TEM''}, it also manifests at low magnifications. As opposed to both BFTEM and dark-field TEM (DFTEM), phase-contrast images are formed when multiple scattered beams are contributing to the formation of the image. Phase-contrast appears in TEM whenever changes occur in the phase of the electron waves scattered from the specimen. In standard STEM mode, phase-contrast primarily arises from shifts in the centre-of-mass (COM) of the convergent beam electron diffraction (CBED) pattern as the electron beam interacts with the sample and its own electric and/or magnetic fields \cite{rose1974phase,lazic2016phase}. Dekkers and de Lang developed the DPC technique in 1974 based on the principles of phase-contrast imaging in STEM as well as deflections on the CBED patterns posed by the specimens \cite{Dekkers1974inventionDPC}. Their method was initially inspired by optical interferometry \cite{franccon1954etude,hamilton1984differential,McCartney1996DPC,amos2003re,lazic2016phase} and it involves measuring the differential intensity of scattered beam signals collected by two opposing sections of a segmented STEM detector. Based on this working principle, Dekkers and de Lang obtained for the first time a ``pure phase image'' in STEM mode without recourse to either defocusing or spherical aberration \cite{Dekkers1977phaseimage}. In 2016, Lazić \textit{et al.} \cite{lazic2016phase} documented the historical development of the DPC technique and identified its primary uses in modern S/TEM \cite{chapman1984investigation,chapman1990modified,shibata2012differential,gauquelin2017determining,yucelen2018phase,de2020resolving,tunes2021irradiation}. These authors emphasised that DPC is still predominantly applied in STEM for studying magnetic samples and to perform atomic-resolution imaging \cite{lazic2016phase}.
	
	\begin{figure}
		\centering
		\includegraphics[width=\linewidth]{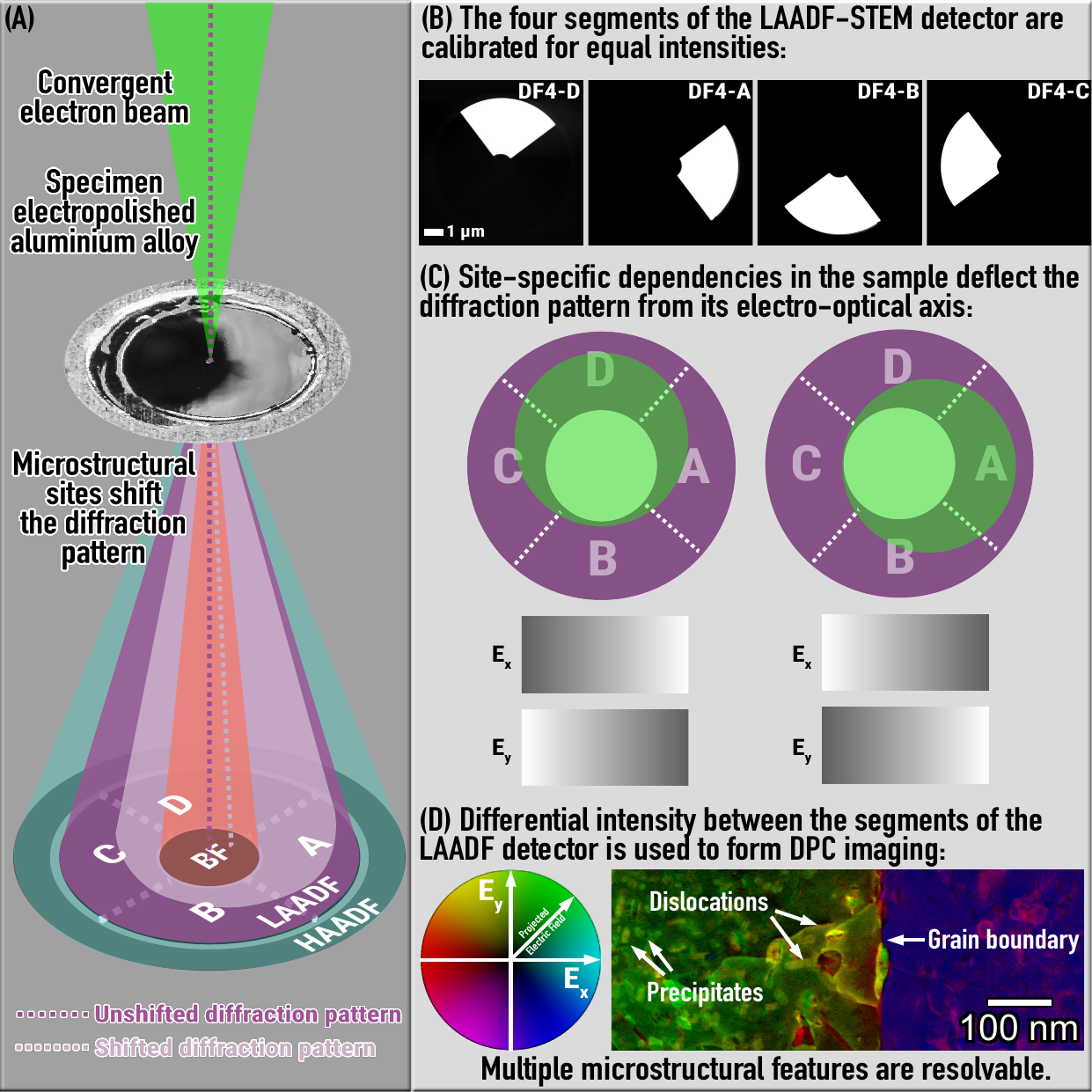}
		\caption{\textbf{Imaging microstructures with DPC} | (A) A convergent electron beam passes through an electron-transparent aluminium alloy sample, where different microstructural sites can rotate the diffraction pattern which is then projected onto a segmented LAADF detector. Before acquiring DPC images, the LAADF-STEM detector (DF4) segments must be both fully covered by the projected beam as well as calibrated to equal intensities, as shown in the actual micrographs in (B). (C) The deflected beam will cause intensity variations across the four LAADF-STEM detector segments. (D) DPC processes the variations between opposing segments and calculate the electron beam direction (0--360\degree) and magnitude (visualised as a gradient from black to one RGB colour in an HSV colour wheel), generating a false-colour micrograph of the alloy microstructure: this ``DPC micrograph'' can be used for nanoscale image segmentation.}
		\label{fig:DPC_principle}
	\end{figure}
	
	A schematic illustrating the principle of DPC-STEM for imaging microstructures is provided in Fig. \ref{fig:DPC_principle}. The electron beam is focused onto an electron-transparent specimen, and the transmitted signal is collected by several detectors, \textit{e.g.}, BF, LAADF, and HAADF as shown in Fig. \ref{fig:DPC_principle}(A). These detectors capture transmitted signals at different angles, which is essential for imaging microstructures in STEM. In this study, our microscope -- Thermo Fisher Talos F200X operating an X-FEG at 200 keV -- has the LAADF detector DF4 is divided into four segments, ordered labelled as DF4-A, DF4-B, DF4-C, and DF4-D. When the STEM probe is active on the specimen, the beam is moved to a nearby hole, then STEM probe alignment is carried out resulting in a centrally aligned and focused beam. Prior to capturing DPC micrographs, the four segments of the DF4 detector are calibrated for equal intensities, as demonstrated in Fig. \ref{fig:DPC_principle}(B). This calibration is achieved using the gain and offset sliders of each DF4 detector segment in the microscope's DPC mode of operation. In this paper, we investigate the hypothesis whether different site-specific microstructural dependencies of aluminium alloys -- such as dislocations, precipitates, grain boundaries and so on -- are able to deflect the CBED pattern from its electro-optical axis. In this way, different segments of the DF4 detector will capture variations in the intensities giving rise to phase imaging as shown in Fig. \ref{fig:DPC_principle}(C). We demonstrate that pure phase imaging of the aluminium alloys' microstructures can be obtained with DPC imaging -- as shown in Fig. \ref{fig:DPC_principle}(D) -- at lower magnifications and that they can be rich in samples' information as much as atomically resolved DPC imaging. Due to the phase colour and intensity nature of DPC imaging, we show in this present research that this technique can be exploited for carrying out microstructural segmentation of complex microstructures at both micro and nanoscale, offering new insights on advanced and complex materials' microstructures. 
	
	It is crucial emphasising that in DPC imaging, as shown in Fig. \ref{fig:DPC_principle}(D), the coloured micrograph represents the local projected electric field of the specimen \cite{de2020resolving}, which is formed from both decomposed x and y components of the electric field as measured from the differential intensity variations in the segments of the DF4 detector.

    The physical origin of phase contrast in DPC imaging can be understood through the phase difference accumulated by the electron beam as it traverses the specimen. For a specimen exhibiting both electrostatic and magnetic properties, the phase difference as shown by McCartney \textit{et al.} \cite{McCartney1996DPC} as:
    \begin{equation}
        \delta \phi = C_E \Delta |V_0| \, t - \frac{e}{\hbar} \iint \mathbf{B} 
        \cdot d\mathbf{A},
        \label{eq:phase_difference}
    \end{equation}
    
    \noindent where $C_E$ is a constant depending on the electron energy, $\Delta |V_0|$ is the difference in mean inner potential, $t$ is the specimen thickness, $\mathbf{B}$ is the magnetic induction, and $\mathbf{A}$ is the planar area enclosed by the electron beam path. In DPC-STEM, the phase gradient is not accessed via beam splitting, but rather through the centre-of-mass shift of the CBED pattern, from which the projected in-plane electric field components are measured experimentally from the differential intensity signals across the four segments of the DF4 detector:
    
    \begin{equation}
        E_x(\mathbf{r}) = -\frac{\partial \phi}{\partial x} \propto 
        \frac{I_A - I_C}{I_A + I_B + I_C + I_D}, \qquad
        E_y(\mathbf{r}) = -\frac{\partial \phi}{\partial y} \propto 
        \frac{I_B - I_D}{I_A + I_B + I_C + I_D},
        \label{eq:DPC_Efield}
    \end{equation}
    
    \noindent where $I_A$, $I_B$, $I_C$, and $I_D$ are the intensities collected by the four DF4 detector segments A--D, and $\mathbf{r} = (x,y)$ is the position vector in the plane of the specimen perpendicular to the incident electron beam. For the aluminium alloy specimens investigated in this study, which are non-magnetic, the second term in Eq.~\ref{eq:phase_difference} vanishes, and the phase contrast measured via Eq.~\ref{eq:DPC_Efield} reflects exclusively the projected electrostatic potential of the microstructure.

    The colour scale for DPC imaging is based on an HSV wheel (hue, saturation, and value) representation:
	
	\begin{itemize}
		\item Hue corresponds to the phase of the specimen's projected electric field, which is spatially represented in the DPC micrograph by a distinct colour.
		\item Saturation indicates the intensity or ``purity'' of the phase signal. In HSV terms, saturation refers to how vibrant or desaturated a colour appears, correlating with the strength or purity of the specimen's projected electric field phase signal.
		\item Value indicates the overall intensity or magnitude of the specimen's projected electric field's, reflecting the brightness of a colour.
	\end{itemize}
	
	In addition to DPC imaging, the segmented LAADF detector can also measure both differentiated DPC (dDPC) \cite{shibata2012differential} and integrated DPC (iDPC) \cite{de2020resolving}. While the first provides a spatial image of a specimen's projected electric charge density, the latter estimates the projected electric potential, which is specially sensitive to light elements such as H \cite{yucelen2018phase,de2020resolving} and O \cite{yucelen2018phase,gauquelin2017determining}. In cases where the specimen exhibit magnetism, either as nanometre-sized precipitates \cite{tunes2021irradiation} or magnetic domain structures \cite{chapman1984investigation,chapman1990modified}, DPC imaging will reflect the samples' projected magnetic field. 
	
	\section{Case studies on advanced aluminium alloys}
	\label{sec:resndis}
	
	\noindent We introduced the principles of the DPC technique in section \ref{sec:DPCprinciple}. This present section details the development of our method and its application of DPC imaging for nanoscale image segmentation of advanced aluminium alloys. We then present five case studies demonstrating its effectiveness in characterising micro- and nanoscale of diverse microstructures. While the present work focuses on aluminium-based alloys as representative case studies, we emphasise that the principles of STEM-DPC imaging and the associated microstructural segmentation workflows are readily transferable to any polycrystalline, multi-phase material system exhibiting local variations in composition, density, or crystal potential. The choice of aluminium alloys serves to demonstrate the versatility of the method across a range of different microstructural landscapes and length scales, yet the approach is equally applicable to steels, titanium alloys, nickel-base superalloys, functional ceramics, and other technologically relevant material classes.
	
	\subsection{Clustering in advanced aluminium alloys}
	\label{sec:resndis:clustering}
		
	Since the late 1930s discovery that aluminium alloys strengthen via GP zone precipitation \cite{guinier1938structure,preston1938structure}, metallurgists are aiming at better understanding of solute-atom aggregation and clustering in the early stages of the precipitate formation \cite{cerny2026early}. Early observations of quenched-in vacancies facilitating solute atom clustering at unexpectedly low temperatures were permitted through advances in X-ray scattering methods between the 1940-1960s \cite{Geisler1948,Kelly1958,Federighi1958,Desorbo1958,Wechsler1958, Rudman1953,Rudman1954,Thomas1959,Tucker1959,Wechsler1959,Berry1959,DeSorbo1959}. In the 1960s up to mid 1970s, the first mechanistic and kinetic models for solute-vacancy binding were developed and tested with pioneering experiments in resistivity and calorimetry, then the phenomenon of clustering (known at the time as ``pre-precipitation'') in aluminium alloys was universally recognised \cite{Chiou1963,Ohta1964,Hashimoto1964,Turnbull1960a,Turnbull1960b,Harrison1960,Ceresara1968,Ceresara1968a,Ceresara1968b,VanTorne1968,Ceresara1969,Kidron1969,Hezel1970,Noble1970,Raman1970,Bowles1973,ElSayed1974}. With the growth of \textit{ab initio} methods and positron annihilation spectroscopy, a clear linkage between clustering and mechanical properties was firstly established in the 1980s \cite{Luiggi1980,Mimault1981,Juhsz1985,Karov1986,Bharathi1988,Carlsson1989,Dutta1989,Nonose1989,DelRio1989}. It was not until the late 1990s that cluster hardening was recognised as a novel strategy for strengthening aluminium alloys \cite{Hono1994,Buchheit1997,Ringer1997,Diak1997,DeBartolo1998,Ringer1998a,Ringer1998b,Zahra1998,Anson1999}. Ringer \textit{et al.} used the emerging technique of APT to visualise solute clustering in AlCuMg alloys, demonstrating that finer clusters enhance mechanical strength before the formation of either GP zones or conventional hardening precipitates \cite{Ringer1997}. Although their findings \cite{Ringer1997} sparked intense debate within the metallurgical community whether these APT-probed clusters should already be classified as GP zones \cite{Zahra1998,Ringer1998a,Ringer1998b}, yet they launched a new era of cluster-hardenable aluminium alloys by correlating APT with mechanical data. 
	
	\begin{figure}
		\centering
		\includegraphics[width=1\linewidth]{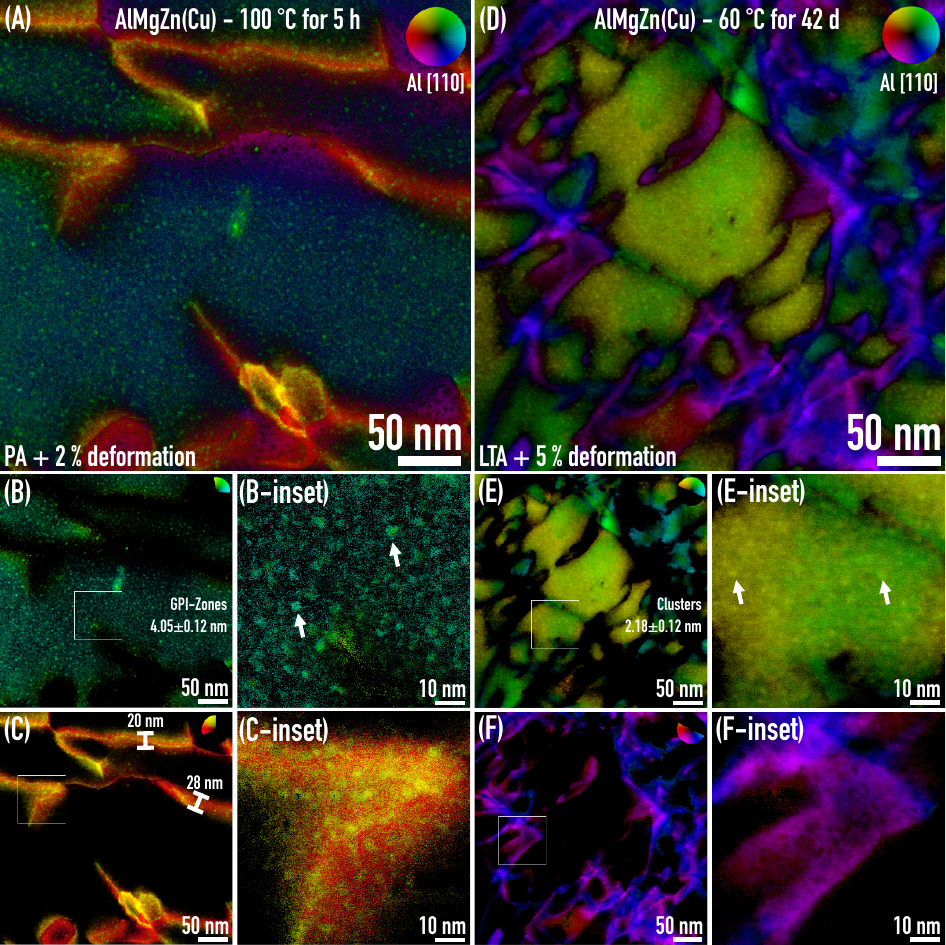}
		\caption{\textbf{Revealing clusters, GPI Zones, dislocations and strain fields in a AlMgZn(Cu) crossover alloy} | The DPC micrograph in (A) shows the microstructure of a pre-aged AlMgZn crossalloy alloy after 2\% plastic deformation. From (A), the segments (B) and (C) were extracted and show in detail, respectively, the alloy matrix with GPI Zones and dislocations with strain fields. Similarly, the DPC micrograph in (D) shows the microstructure of a long-term-aged AlMgZn crossover alloy after 5\% plastic deformation. From (D), the segments (E) and (F) were extracted and show in detail, respectively, the alloy matrix mainly with small clusters and dislocations. As expanded in section \ref{sec:resndis:clustering}, direct quantitative and qualitative comparisons between both alloys' microstructures are permitted with the DPC technique. The insets in the segmented images (B), (C), (E), and (F) show enlarged areas of their respective micrographs.}
		\label{fig:clusters}
	\end{figure}
	
	Clustering has become a key materials design tool in modern lightweight metallurgy, yet significant challenges remain. Dumitraschkewitz \textit{et al.} \cite{dumitraschkewitz2018clustering} and de Vaucorbeil \textit{et al.} \cite{deVaucorbeil2018} recently reviewed the phenomenon of clustering in aluminium alloys, highlighting that key challenges remain in detection, characterisation, and in the own definition of clusters as distinct physical entities. APT is still the major experimental method for detecting clusters, but its inherent high-intensity electric field is known to cause surface migration of mobile atoms, which complicates the accuracy and reliability of the APT's signal-reconstruction. Although some recent APT experiments have ruled out natural ageing in aluminium alloys when the samples' dimensions lie on the nanoscale \cite{dumitraschkewitz2019size} -- which could potentially impact cluster detection and analysis with APT -- distinguishing true clusters from both random composition fluctuations or even smaller GP zones still requires rigorous statistical methods and reliable cluster-finding algorithms \cite{dumitraschkewitz2018clustering,deVaucorbeil2018}. Recent work correlate the atomic-scale details of cluster composition to macroscale properties, specially mechanical behaviour \cite{Wang2024,aster2023strain,aster2024unraveling,aster2025clustering}. These studies suggested that plastic deformation and novel heat treatments can enhance the population of strain-induced vacancies, accelerating clustering and improving the mechanical properties of new aluminium alloys. Some of these cluster-hardenable aluminium alloys have demonstrated mechanical properties comparable to, or even rivalling, those of peak-aged commercial aluminium alloys  \cite{Wang2024,aster2023strain,aster2024unraveling,aster2025clustering}. This highlights the potential of clustering research to further enhance the development of cluster-hardened aluminium alloys, offering an alternative strengthening mechanism beyond conventional ageing treatments, which impact future industrial developments.
	
	Here we present a case-study on the use of phase-contrast imaging via STEM-DPC for directly observing early-stage precipitates in aluminium alloys. For that, we screened the microstructure of an aluminium crossover alloy -- an AlMgZn(Cu) alloy that bridges the 5xxx and 7xxx series alloys \cite{aster2025clustering,stemper2021giant,stemper2022potential} -- subjected to two different heat-treatments: (i) Pre-Ageing (PA) at 100ºC for 5 hours after 2\% plastic deformation and (ii) Long-Term Ageing (LTA) at 60ºC for 42 days after 5\% plastic deformation. The DPC micrograph in Fig. \ref{fig:clusters}(A) shows the microstructure of the alloy subjected to PA after 2\% deformation whereas Fig. \ref{fig:clusters}(D) shows the microstructure of the same alloy subjected to LTA after 5\% deformation. Matrix, precipitates and dislocations are visible in both DPC micrographs in Figs. \ref{fig:clusters}(A) and \ref{fig:clusters}(B), demonstrating that DPC imaging can be used to differentiate site-specific microstructural dependencies in advanced aluminium alloys.
	
	From DPC imaging, nanoscale image segmentation is carried out by selecting the hue interval corresponding to phase (or the phase range) of the transmitted electron beam scanning through precipitate site-specific positions in the microstructure. As shown in Fig. \ref{fig:clusters}(B), precipitates -- previously identified in this crossover alloy as GPI zones \cite{stemper2021giant} -- can be effectively segmented from the DPC micrograph in Fig. \ref{fig:clusters}(A). Similarly, segmentation of precipitates in the LTA alloy is demonstrated in Fig. \ref{fig:clusters}(E). A detailed analysis of these phase segmentations (Figs. \ref{fig:clusters}(B-inset) and \ref{fig:clusters}(E-inset)) allows direct comparison of the effects of PA and LTA treatments on the aluminium crossover alloy. In the PA-treated sample, precipitates measure 4.05$\pm$0.12 nm in diameter, whereas in the LTA-treated sample, they are approximately 50\% smaller at 2.18$\pm$0.12 nm. In the PA case, GPI zones are clearly distinguishable from the matrix, while in the LTA case, the smaller precipitates -- considered clusters in \cite{aster2023strain,aster2024unraveling,aster2025clustering} -- appear slightly out-of-phase with the matrix. These results demonstrate that DPC imaging can detect early-stage precipitates as small as ~2 nm. As they evolve into larger GP zones, their phase separation from the matrix becomes more pronounced. Moreover, DPC imaging combined with nanoscale image segmentation enables the quantification of numerous clusters within a single field of view and from a micrograph that took not long than $\sim$1 minute for registering. Unlike prolonged APT experiments, which require post-processing with cluster-finding algorithms to identify clusters, DPC enables direct imaging of clusters in aluminium alloys, capturing their size, morphology, and distribution within the matrix.
	
	DPC micrographs can be used for segmenting not only clusters and GPI zones, but also dislocations and their associated strain fields, which are also responsive to DPC imaging. While strain fields of dislocations have been detected and quantified using other phase-contrast and diffraction methods, such as diffracted beam interferometry (DBI) \cite{herring2021phase} and 4D-STEM \cite{ophus2019four,mahr2021accurate}, this study demonstrate the use of DPC imaging for qualitative microstructural mapping of both dislocations and their strain fields. As shown in Figs. \ref{fig:clusters}(C) and \ref{fig:clusters}(F), nanoscale segmentation reveals dislocations in both PA and LTA deformed alloys. The key difference between the two cases is the degree of deformation: the PA alloy underwent 2\% deformation, while the LTA alloy experienced 5\%. From extracted segments in Figs. \ref{fig:clusters}(C) and \ref{fig:clusters}(F), we estimated with ImageJ an areal fraction corresponding to dislocations segments and their strain fields of approximately 29\% and 45\%, respectively for the PA and LTA alloys, confirming that more plastic deformation yields more dislocations in the latter. In the PA case, the lower areal density of dislocations allows their strain fields to be more distinctly visualised, extending to approximately 20–28 nm from the dislocation core. In contrast, the slightly higher deformation in the LTA case leads to dislocation accumulation and interaction, forming a dislocation network where strain fields become entangled, but remain detectable. When focusing to the dislocations, Fig. \ref{fig:clusters}(C-inset) shows that GPI zones accumulate within both the strain field and core regions of the dislocations in the PA case. In the LTA alloy case, clusters are neither visible within the dislocation core nor around the strain fields as shown in Fig. \ref{fig:clusters}(F-inset). We suggest that our nanoscale segmentation method via DPC could be also applied for studying dislocation-precipitates interaction. 
	
	Both GP zones and clusters are of different chemical composition and local crystallographic order when compared to the aluminium matrix, thus shifts of the phase of the incident electron beam can be attributed to different projected electric potentials arising from the beam-sample local interaction. Dislocations, on the other hand, do not bear differences in chemical composition with the matrix. Their phase shift is purely attributed to rotation of the convergent electron beam diffraction pattern at both dislocation core and strain field \cite{mcmorran2011electron,herring2011new,herring2021phase}. In materials science, the Burgers circuit of a dislocation completes a 360${\degree}$ rotation around its core. This induces a 2$\pi$ phase shift to the incident electron beam at the dislocation position. At the centre of a 2$\pi$ phase shift within a dislocation position lies a quantum mechanical singularity in space (known as ``dislon'' \cite{li2018theory}) where the phase is not well defined, causing beam interference effects that, according to McMorran \textit{et al.} \cite{mcmorran2011electron}, can twist the electron beam, producing a vortex. In our present work, the DPC imaging technique has not revealed a 2$\pi$ phase shift at the dislocation positions as McMorran \textit{et al.} \cite{mcmorran2011electron} and other authors previously reported \cite{penkova2019dislocation,herring2021phase}. Herring \cite{herring2021phase} also did not observe such 2$\pi$ phase shift applying the DBI technique on pure aluminium foils, attributing this fact to possibility that the Burgers vector and habit plane of these dislocations do not align to reveal the 2$\pi$ phase shift. Alternatively, both strain relaxation due to the thin film effect and the formation of dislocation networks may prevent the completion of the 2$\pi$ phase shift. Interestingly, in our aluminium crossover alloys presented in Fig. \ref{fig:clusters}, with the nanoscale image segmentation technique, we show that the phase shift between the dislocation core and strain field in the PA alloy with 2\% deformation is around $\pi/4$ whereas for the LTA alloy with 5\% deformation, the phase difference between the dislocation core and the strain field is approximately $\pi/2$. This indicates the electron beam has a tendency to shift towards a complete 2$\pi$ phase shift starting from the vicinities of the strain field to the dislocation core, but dislocation number density, interaction and entanglement, and deformation level may play a role that deserves further investigation.
	
	\subsection{Paint-bake response of novel aluminium alloys}
	\label{sec:resndis:PB_2pc_AlMgZn_crossalloy}
	
	\noindent The paint-bake (PB) cycle is a thermal treatment process integral to automotive manufacturing, in which body panels are heated to approximately 170--190°C for 20--30 minutes in total. This thermal exposure is intended for paint curing, but it is also exploited to promote precipitation in aluminium alloy sheets: a phenomenon known as the paint-bake response or bake-hardening, which was primarily invented for steels \cite{Okamoto1989paintbake}, but later evolved to aluminium alloys \cite{Fujita1995paintbake,stemper2021giant}. As aluminium has become increasingly prevalent in automotive body structures owing to its favourable strength-to-weight ratio \cite{Ota2020trends, Scharifi2023forming}, optimising the bake-hardening response of sheet alloys has attracted considerable research interest \cite{Keerthipalli2023review, Rolle2022paintbake}. In the 6xxx (Al--Mg--Si) alloy series, hardening mechanism is primarily driven by the formation of the \textbeta$''$-phase, but GP zone precursors are also reported to form during baking \cite{Uggowitzer2001paintbake, Rometsch2012paintbake, Gong2021giantbake, Yeh2024paintbake}. The PB response is also sensitive to prior natural ageing, which can suppress subsequent artificial ageing kinetics. Pre-ageing treatments have been developed to mitigate this effect \cite{Rometsch2012paintbake, Yeh2024paintbake, stemper2021giant}. Similar bake-hardening behaviour has been reported in Al--Mg--Zn alloys \cite{Balderach2003paintbake}, while the higher-strength 7xxx series presents additional complexity as the existing temper microstructures may evolve to different equilibrium states during the paint-bake thermal cycle \cite{Harrison2017paintbake, Ma2023paintbake}.
	
	\begin{figure}
		\centering
		\includegraphics[width=\linewidth]{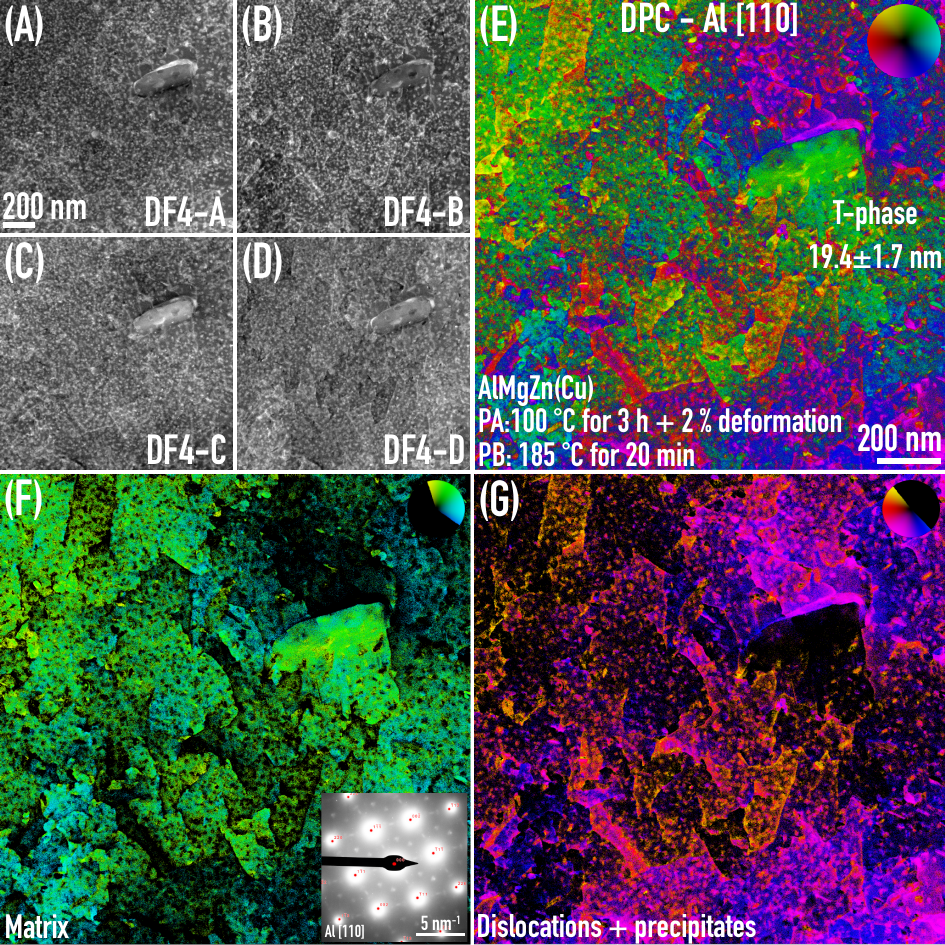}
		\caption{\textbf{Mosaic of intermediate T$^{\prime}$-phase precipitates and dislocations in a paint-baked alloy} | The  AlMgZn(Cu) crossover alloy was subjected to paint-bake heat treatment and 2\% deformation. The LAADF micrographs from (A) to (D) show the segmented LAADF signal on the DF4 detector segments. With the signals from (A) to (D), a DPC micrograph is reconstructed and shown in (E) revealing dislocation loops and precipitates. With nanoscale image segmentation, the DPC micrograph in (E) can be decomposed into matrix (F), dislocation and precipitates (G). Notes: The scale bar in (A) applies to (B-D) and the scale bar in (E) also applies to (F) and (G). The inset in (F) shows the SAED pattern indexed to be the Al [110] zone axis.}
		\label{fig:PB_2pc_AlMgZn_crossalloy}
	\end{figure}
	
	Stemper \textit{et al.} \cite{stemper2021giant} recently demonstrated a thermomechanical processing concept for new AlMgZn(Cu) crossover alloys -- based on commercial AA5182 modified with Zn and Zn+Cu additions -- that yields an exceptionally large hardening response during a short paint-bake treatment (20 min at 185°C). Following a 3-hour pre-ageing step at 100°C, both alloys form a high number density of GPI zones and exhibit favourable formability due to enhanced work hardening. Cu additions further increase the GPI zone number density and size by stabilising early-stage clusters. Without prior deformation, the paint-bake response is modest in the Cu-free alloy yielding 47 MPa, but reaching 127 MPa with Cu: this hardening effect was attributed to the nucleation of intermediate T$^{\prime}$-phase precipitates from the pre-existing GPI zones. Critically, Stemper \textit{et al.} discovered that applying a 2\% pre-strain between the pre-ageing and paint-bake stages introduces dislocations that act as preferential nucleation sites and accelerate solute transport via pipe diffusion, boosting the total hardening response to a reported maximum of 184 MPa and a final yield strength of 410 MPa in the alloy with Cu -- exceeding the paint-bake response of conventional automotive AlMgSi alloys, demonstrating that new crossover alloys can be a viable route toward a single-alloy automotive sheet concept with both high in-service strength and good formability during forming.
	
	We applied the technique of STEM-DPC to investigate the intricate microstructure of the new AlMgZn(Cu) crossover alloy with 2\% pre-strain between the pre-ageing and paint-bake treatments. These results are shown in Fig. \ref{fig:PB_2pc_AlMgZn_crossalloy}. The micrographs in Fig. \ref{fig:PB_2pc_AlMgZn_crossalloy}(A--D) show the four segments of the LAADF detector which are used to compose the DPC micrograph in Fig. \ref{fig:PB_2pc_AlMgZn_crossalloy}(E): a mosaic of colours reveals a matrix full of dislocation segments and small nanometre-sized precipitates, identified by Stemper \textit{et al.} as intermediate T$^{\prime}$-phase. From the DPC image in Fig. \ref{fig:PB_2pc_AlMgZn_crossalloy}(E), image segmentation can be performed based on the colour-phase space of the transmitted electron beam after interaction with the sample: Fig. \ref{fig:PB_2pc_AlMgZn_crossalloy}(F) shows only the matrix regions of the alloy whilst in Fig. \ref{fig:PB_2pc_AlMgZn_crossalloy}(G) both dislocation segments and the hardening precipitates can be seen. The selected-area electron diffraction (SAED) pattern inset in Fig. \ref{fig:PB_2pc_AlMgZn_crossalloy}(F) points to a single type of hardening precipitates present in the microstructure of the alloy after pre-ageing, pre-straining and paint-bake -- consistent with the previous observation of T$^{\prime}$-phase reported in a recent work of Stemper \textit{et al.} \cite{stemper2021giant}. This segmentation analysis visually indicates that the precipitates tend to agglomerate within the dislocation segments -- clearly seen in the segmented micrograph shown in Fig. \ref{fig:PB_2pc_AlMgZn_crossalloy}(G). This observation validates the idea that pre-ageing/pre-straining accelerate the hardening response due to dislocations trapping solutes and serving as heterogeneous sites for precipitation -- a phenomenon known as giant PB hardening recently reported by Stemper \textit{et al.} in new crossover aluminium alloys \cite{stemper2021giant,stemper2022potential}. 
	
	\subsection{Overageing effects in aerospace aluminium alloys (AlZnMg)}
	\label{sec:resndis:overaged7075}
	
	\noindent Overageing heat treatments have high industrial significance in AA7xxx series aluminium alloys, which are widely employed in aerospace and structural applications, where both mechanical performance and environmental resilience are critical. Whilst the peak-aged T6 temper maximises strength, it renders the alloy highly susceptible to stress corrosion cracking (SCC) and pitting corrosion. Overageing to T73 or T7451 temper is historically reported to substantially improve corrosion and SCC resistance, although a reduction in mechanical properties -- plasticity in particular -- is consequently reported \cite{Doig1977, Viana1999, Dey2025}. This trade-off has driven significant research into optimising heat treatment schedules, including retrogression and re-ageing (RRA) approaches that seek to recover T6-level strength whilst retaining the favourable grain boundary microstructure associated with overaged tempers \cite{Viana1999}. Beyond corrosion performance, overageing has also been shown to provide meaningful relief of residual stresses in thick-section components -- an important empirical consideration in aerospace manufacture \cite{Robinson2017}. The sensitivity of these properties to even minor thermal excursions further underscores the need for precise thermal processing control in industrial practice \cite{Paglia2007}.
	
    Ryggetangen \textit{et al.} \cite{Ryggetangen2025sped} demonstrated that the microstructures of AlZnMg alloys aged for 264 h at 120 °C are composed of up to six different hardening precipitate phases per alloy (T$'$ and \texteta -type phases).  They applied the emerging technique of scanning precession electron diffraction (SPED), a variant of scanning nanobeam diffraction in which the electron beam in STEM-mode is given a deliberate conical motion as it is stepped across the sample. A diffraction pattern is collected at every beam position, producing a four-dimensional dataset spanning both real and reciprocal space. The precession motion serves two purposes: it integrates a greater number of diffraction reflections at higher scattering angles, and it reduces the dynamical diffraction artefacts that complicate conventional selected-area diffraction. The technique is particularly advantageous for aluminium alloys, where slight foil bending would otherwise introduce orientation artefacts, and its spatial resolution is sufficient to resolve individual nanoscale precipitates across regions of approximately \textmu m$^2$.
	
	\begin{figure}
		\centering
		\includegraphics[width=\linewidth]{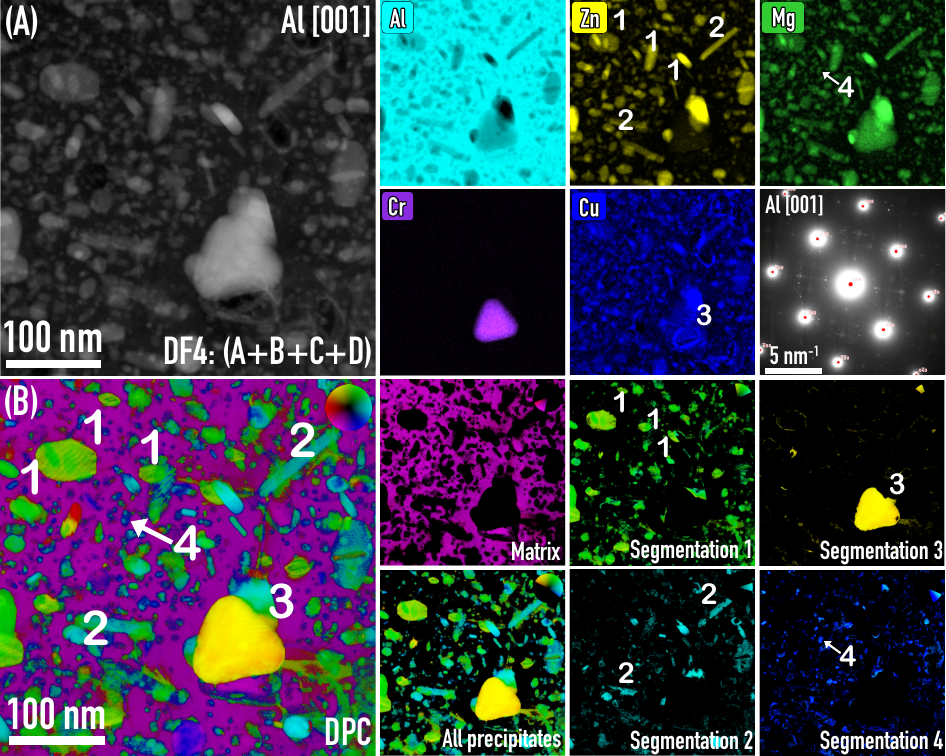}
		\caption{\textbf{Correlative microscopy with STEM-EDX and -DPC of overaged microstructures} | AA7075-T7 is a high-strength aerospace alloy in the overaged state. (A) Presents a composite LAADF micrograph with all the segments of the DF4 detector, STEM-EDX elemental maps for Al, Zr, Mg, Cr, Cu, and the SAED pattern of the grain taken along the [001] zone axis. (B) Shows the complementary DPC analysis from the same region in (A) where 4 segmented images of distinct precipitates could be extracted from the DPC micrograph along with the alloy's matrix.}
		\label{fig:overaged7075}
	\end{figure}
	
	We applied our STEM-DPC-based microstructural segmentation technique to an AA7075 alloy in the T7 (overaged) temper, and the results are given in Fig. \ref{fig:overaged7075}. The combined four-segment-micrographs from the DF4 detector are shown in the overlay in Fig. \ref{fig:overaged7075}(A). We report herein correlative STEM-EDX and -DPC microscopy in this exact region-of-interest (ROI). The set of STEM-EDX maps also present in Fig. \ref{fig:overaged7075}(A) shows that the overaged AA7075-T7 presents a microstructure composed of an aluminium matrix with: (i) Zn-Mg-rich nano-precipitates with both needle-like and rounded-like morphologies; (ii) a Cr-rich large dispersoid phase; (iii) and areas where Cu seems to be enriching the Zn-Mg precipitates in different concentrations. In qualitative STEM-EDX maps, colour saturation directly corresponds to elemental concentration. Variations in Zn, Mg, and Cu saturation at site-specific locations suggest the presence of multiple hardening phase variants -- analogous to the findings of Ryggetangen \textit{et al.} \cite{Ryggetangen2025sped}. The SAED pattern in Fig. \ref{fig:overaged7075}(A) also corroborates the presence of multiple phases beyond the aluminium matrix.
	
	The STEM-DPC and its corresponding nanoscale-image segmentation are shown in Fig. \ref{fig:overaged7075}(B). The technique allows full segmentation of the matrix from all the precipitates within the alloy. Segmentations 1, 2, 3, and 4 correspond to the precipitate variants found in the ROI, which were segmented by the DPC signal/colour -- a measure of the projected electric field on the sample. Segmentation 3 corresponds mainly to the Cr-rich dispersoid phase. Segmentations 1 and 2 indicate the presence of morphologically distinct precipitates with different phase fractions -- one rounded with likely higher phase fraction, the other needle-like with apparently lower phase fraction. Their differing DPC signals reflect variations in both chemical composition and, possibly, crystal structure (although DPC cannot directly attest crystallography unlike SPED), resulting in distinct projected electric fields at each site. These two segments are therefore attributed to different variants of Zn–Mg-rich precipitates. Segmentation 4 corresponds to the smallest precipitate identified in the overaged alloy and is interpreted as a precursor phase that has not yet undergone the transformation and coarsening observed in segmentations 1 and 2. We hypothesize that segmentations 1 and 2 correspond to \texteta\ and T, while segmentation 4 corresponds to precursors of \texteta, or T$'$. If the DPC signal at both micro- and nanoscales were further calibrated against SPED and 4D-STEM methods, it could become a powerful and efficient technique for characterising the microstructures of age-hardenable metallic alloys. A typical DPC micrograph, such as that shown in Fig. \ref{fig:overaged7075}(B), can be acquired in approximately 10--30 seconds at 2k resolution.
	
	\subsection{Anti-corrosion strategies in aerospace aluminium alloys (AlCuMg)}
	\label{sec:resndis:anodic2024}
	
	\noindent Anodisation is an electrochemical technique to grow a protective oxide layer onto aluminium alloys that can serve two major purposes: (i) corrosion protection and (ii) aesthetics. The resilience of such anodic layers is intrinsically related to their microstructure, which has already been addressed by multiple electron-microscopy characterisation studies in early 1950s \cite{Keller1953,Renshaw1961,Hoar1963,Franklin1963,Csokn1964,Paolini1965}. The seminal SEM work of Keller \textit{et al.} \cite{Keller1953} at the former Aluminium Company of America (nowadays known as ALCOA corporation) has unravelled that the anodic coatings grow following a hexagonal pattern containing a single-pore structure. The pore size (10--30 nm), shape and morphology are fully dependent on the choice of electrolyte, electric potential, and time, among other parameters. Recent research has revisited strategies for growing anodic oxide layers specifically on aerospace-grade aluminium alloys, driven by the need to replace Cr-based industrial coatings, which are carcinogenic \cite{Carlton2003}. Anodic layers with pores filled with metal nanoparticles \cite{ramirez2020tartaric,ramirez2022corrosion} or sealed with sol-gel \cite{ramirez2023selfheal} are examples of approaches used to enhance the corrosion resistance of anodic films in aerospace aluminium alloys.
	
	\begin{figure}
		\centering
		\includegraphics[width=0.9\linewidth]{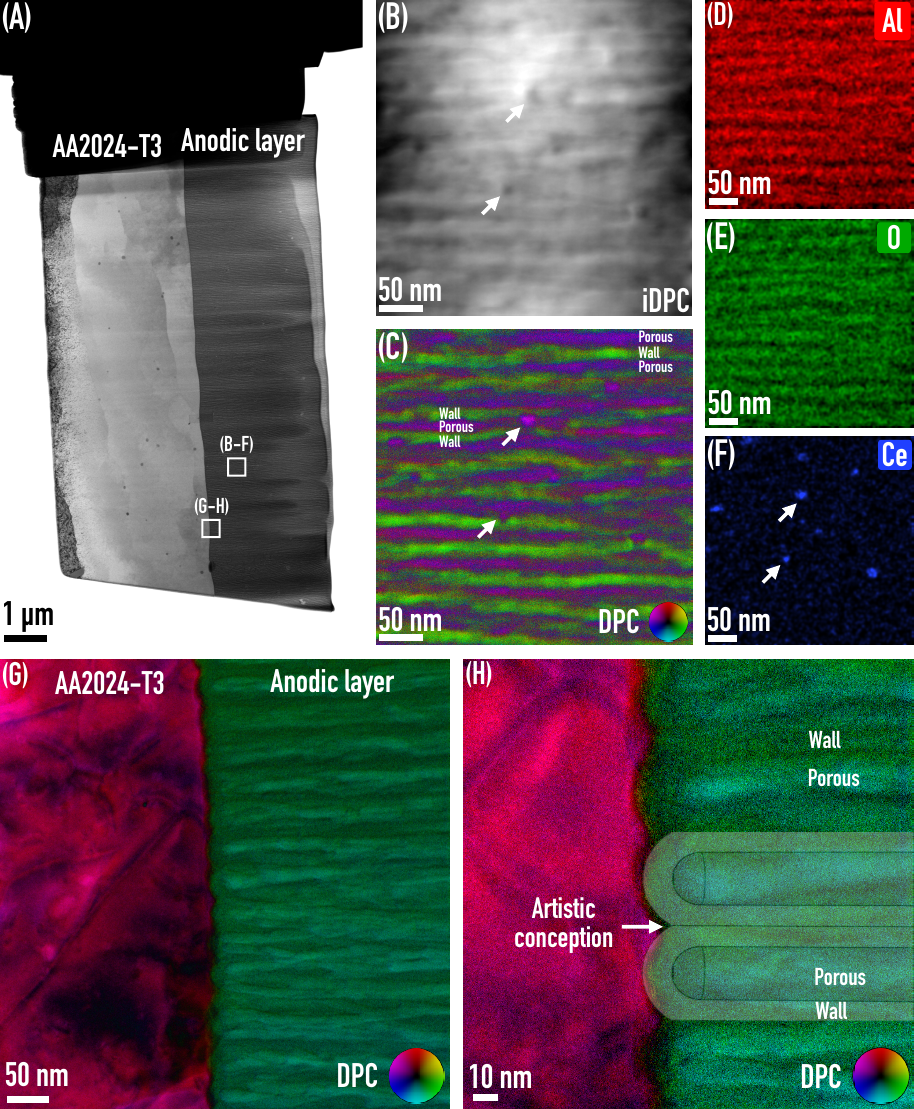}
		\caption{\textbf{Advancing aluminium alloys' corrosion protection research with DPC} | BF-STEM micrograph in (A) shows the sample's cross-section of an AA2024-T3 alloy substrate with a grown anodic layer. Micrographs (B) and (C) show both the iDPC and DPC micrographs from the anodic layer, where anodic walls, pores and Ce nanoparticles are distinguishable. Complementary STEM-EDX elemental mapping in (D-F) reveals aluminium oxide nature of the anodic layer as well as the presence of Ce nanoparticles. DPC micrographs (G) and (H) taken at the interface between the alloy and the anodic layer revealing in nanoscale detail the ``nanotubular'' morphological aspect of the latter, matching with the overlaid artistic conception by Prada-Ramirez \textit{et al.} \cite{ramirez2020tartaric}.}
		\label{fig:anodic2024}
	\end{figure}
	
	STEM-DPC was applied here to examine an anodic layer grown on an aerospace-grade high-strength aluminium alloy, clad AA2024-T3 (Al--Cu). The anodic layer was impregnated with Ce nanoparticles, which have been shown to significantly enhance the corrosion resistance of the alloy \cite{ramirez2020tartaric}. The FIB lamella in Fig. \ref{fig:anodic2024}(A) shows both the alloy substrate and the anodic layer under BF-STEM. STEM-DPC imaging of the anodic layer reveals that both the nanoporous structure and the Ce nanoparticles are detectable via phase contrast. The iDPC micrograph in Fig. \ref{fig:anodic2024}(B) maps the projected electric potential across the anodic layer, in which the metal nanoparticles are clearly visible. The DPC micrograph in Fig. \ref{fig:anodic2024}(C) enables the oxide pore walls and the Ce nanoparticles to be distinguished. A complementary set of STEM-EDX maps in Figs. \ref{fig:anodic2024}(D--F) corroborates the presence of Ce nanoparticles within the tubular aluminium oxide layer.
	
	The clad AA2024-T3 substrate and the anodic layer can be seen in the DPC micrograph in Fig. \ref{fig:anodic2024}(G). A higher magnification assessment in Fig. \ref{fig:anodic2024}(H) shows that DPC allows precise detection of the porous regions within the tubular structure: the artistic conception overlay, drawn from the seminal work of Keller \textit{et al.} \cite{Keller1953}, reveals that DPC is, perhaps, the first technique to image the entire morphology of an anodic layer with a contrast that allows direct identification of site-specific features: wall, porous regions and nanoparticles filling the pores. We anticipate that STEM-DPC will emerge as a powerful technique for microstructural characterisation in the context of electrochemistry and anodisation strategies of functional alloys.

	\subsection{Grain analysis in nanocrystalline aluminium thin films}
	\label{sec:resndis:grains}
	
	\noindent Grain structure, grain boundaries, and grain size distributions are fundamental microstructural parameters governing a wide range of physical properties in metallic materials, including electrical conductivity and mechanical ductility \cite{Meyers2006, Orlova2016, Khatamsaz2025}. To investigate grain size dependencies at the nanoscale, nanocrystalline thin films offer a tractable model system: they can be deposited by magnetron sputtering and subsequently examined by \textit{in situ} TEM heating or electrical biasing experiments. Pure aluminium films of 100 nm thickness were produced by magnetron sputtering directly onto MEMS chips with the Protochips Fusion \textit{in situ} holder \cite{Protochips2020}. These chips incorporate a 50 nm amorphous silicon nitride (SiN) membrane to ensure electron transparency and to serve as support for the deposited film. As this membrane is made with a refractory ceramics, it is inert with respect to the deposited aluminium. The resulting polycrystalline thin films are well suited to \textit{in situ} investigations of thermally or electrically driven phenomena such as grain growth, grain rotation, and precipitate formation in alloys.
	
		\begin{figure}[hb!]
		\centering
		\includegraphics[width=1\linewidth]{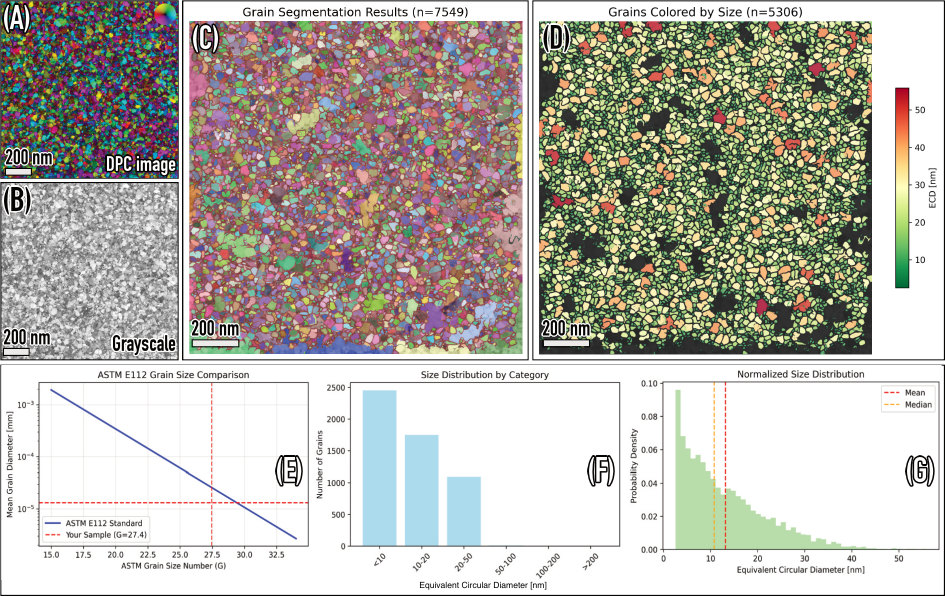}
		\caption{\textbf{Grain segmentation and ASTM E112 standard grain analysis via STEM-DPC} | (A) Shows a typical DPC microstructure obtained from nanocrystalline thin films. The micrograph in (B) shows the same micrograph/region exported with grayscale contrast (a.k.a. log10 in Thermo Fisher Velox). The grain segmented micrograph in (C) shows the result application of convolutional neural network (\textit{i.e.} U-Net) in micrograph (B): the algorithm was able to identify the grain boundaries with the current nanoscale spatial resolution. Further screening of the U-Net is shown in micrograph (D) where equivalent circular diameter (ECD) analysis allows to plot different grain-size ranges after outlier removal. From the U-Net analysis coupled with STEM-DPC, (E) the ASTM grain size number, (F) size distribution diagram, and (G) an histogram of ECDs is permitted.}
		\label{fig:hasi}
	\end{figure}

	Tracking these microstructural changes in conventional TEM mode is complicated by bending contours, which obscure grain boundaries and interior features. In STEM mode, grain boundaries of nanocrystalline films on amorphous substrates often exhibit poor contrast, particularly when grain dimensions fall below 100 nm. DPC imaging mitigates these limitations by exploiting the additional phase information encoded in the segmented DF4 detector signal: individual grains generate different phase shifts, due to their distinct crystallographic orientations, and they therefore appear as distinct colours in the DPC micrograph, making grain boundaries directly visible and facilitating their delineation.
	
	This improved contrast directly benefits automated image segmentation. Established grain segmentation algorithms developed for conventional TEM and STEM suffer from reduced success rates in nanocrystalline films due to the contrast artefacts described above \cite{Barmak2024, Xu2024}. Here, a purpose-built segmentation pipeline was applied to the DPC micrographs, combining a trained U-Net convolutional neural network \cite{Matthews} with the Segment Anything Model (SAM)\cite{Kirillov2023} using the ASTM Standard E112 for average grain size determination \cite{E112}. The pipeline was originally developed for SEM micrographs \cite{DigitalSreeni} and adapted here for DPC input; notably, the U-Net alone -- without the Segment Anything Model -- proved sufficient for grain identification under these imaging conditions.
	
	The DPC micrograph of a 100 nm thick pure aluminium film is presented in Fig. \ref{fig:hasi}. The corresponding greyscale image -- Fig. \ref{fig:hasi}(B) -- derived from the DPC magnitude channel, was used as input to the segmentation algorithm. Identified grains are categorised by equivalent circular diameter into four size classes, as shown in Fig. \ref{fig:hasi}(C): 2454 grains below 10 nm, 1749 between 10 and 20 nm, 1095 between 20 and 50 nm, and only 8 between 50 and 100 nm. Regions of insufficient contrast, where the DPC signal was too weak for reliable boundary detection, could not be segmented and appear as black areas in Fig. \ref{fig:hasi}(C). A continuous colour map in Fig. \ref{fig:hasi}(C) encodes grain size from the smallest (dark green) to the largest (dark red), providing an immediate visual impression of the size distribution across the field of view. The U-Net identified 7548 grains in total; following outlier removal using the criterion $Q_3 + 3 \cdot \mathrm{IQR}$, 5306 grains were retained for statistical analysis. The equivalent circular diameter ranges from 2.5 to 55.2 nm, yielding a ratio of 21.7 between the largest and smallest grains. The size distribution has a mean of 13.2 nm, a median of 10.8 nm, and a standard deviation of 8.99 nm, consistent with a right-skewed distribution dominated by sub-20 nm grains. The segmentation output is compiled into a summary report providing grain-level statistics including equivalent circular diameter, aspect ratio, and circularity for every detected grain.

	\section{Conclusion and future works}
	\label{sec:conclusion}
	
	\noindent This work has demonstrated the application of STEM-DPC imaging as a versatile technique for nanoscale image segmentation of advanced aluminium alloys. By exploiting the phase-contrast sensitivity of a segmented LAADF detector (DF4), the projected electric field of the specimen is encoded as a false-colour micrograph whose hue--saturation--value space can be decomposed 
	into independent microstructural segments. The following principal conclusions are drawn from the five case studies presented:
	
	\begin{enumerate}
		
		\item In the deformed AlMgZn(Cu) crossover alloy, DPC imaging resolved nanoclusters as small as $\approx$2\,nm and GPI zones, demonstrating that phase separation from the aluminium matrix becomes increasingly pronounced as clusters evolve into GP zones. Nanoscale segmentation further enabled the direct visualisation and quantification of dislocation strain fields, extending approximately 20--28\,nm from dislocation cores, and revealed preferential accumulation of GPI zones within dislocation cores and their	surrounding strain fields in the pre-aged condition.
		
		\item In the paint-baked AlMgZn(Cu) crossover alloy subjected to 2\% pre-strain, DPC-based segmentation unambiguously resolved intermediate T$'$-phase precipitates co-located along dislocation segments, providing	direct visual evidence for dislocation-assisted heterogeneous nucleation as the origin of the giant paint-bake hardening response reported by Stemper \textit{et al.} \cite{stemper2021giant,stemper2022potential}.
		
		\item Applied to overaged AA7075-T7, DPC segmentation resolved at least four distinct precipitate variants within a single region of interest,	distinguished by their projected electric field phase signal. This multiplicity of hardening phases in the overaged condition is consistent with the recent findings of Ryggetangen \textit{et al.} \cite{Ryggetangen2025sped}. When combined with correlative STEM-EDX 
		mapping, DPC-based segmentation provides a time-efficient route for assessing the thermodynamic equilibrium microstructure of age-hardenable aluminium alloys, complementing and accelerating characterisation strategies that would otherwise rely exclusively on more complex or time-intensive diffraction-based techniques.
		
		\item In the anodic corrosion-protective coating grown on AA2024-T3, DPC and iDPC imaging resolved the nanoporous oxide wall structure and cerium nanoparticle distribution within the tubular anodic layer, establishing DPC as a high-contrast technique for site-specific characterisation of electrochemically grown functional coatings.
		
		\item Grain boundary delineation in nanocrystalline pure aluminium thin films was demonstrated via DPC imaging coupled with a U-Net convolutional neural network segmentation pipeline, yielding grain statistics --- equivalent circular diameter, aspect ratio, and circularity --- consistent with ASTM Standard E112 requirements. The approach provides a practical route to automated, statistically significant grain size analysis at spatial scales inaccessible to conventional STEM contrast mechanisms.
		
	\end{enumerate}
	
	Collectively, these results establish STEM-DPC as a powerful and efficient technique for multimodal microstructural characterisation of aluminium alloys across micro- and nanoscales. A typical DPC micrograph can be acquired in 10--30 seconds at 2k resolution, comparing favourably with the extended acquisition and post-processing times required by APT, SPED, or 4D-STEM. Future work coupling DPC with 4D-STEM and expanding its quantitative calibration against diffraction-based phase identification methods will further consolidate its role as an indispensable tool in modern nanometallurgy \cite{allioux2020bi,coradini2024unravelling,raabe2022making}.

    While the case studies presented herein focus on aluminium-based alloys, the underlying principles of STEM-DPC imaging and microstructural segmentation are by no means restricted to a single material class. Any polycrystalline, multi-phase system that exhibits local variations in composition, density, or crystal potential is, in principle, amenable to DPC characterisation. We therefore anticipate that the workflows demonstrated here can be transferred readily to steels, titanium alloys, nickel-base superalloys, functional ceramics, and other technologically relevant material systems, thereby establishing STEM-DPC segmentation as a versatile and widely accessible tool in the modern materials scientist's characterisation portfolio.
	
	
	\section{Experimental Section}
	\label{sec:matnmet}
	
	\subsection{Aluminium alloys used in this work}
	\label{sec:matnmet:AAs}
	
	\subsubsection{Cluster-hardenable AlMgZn crossover alloy} 
    \noindent The cluster-hardenable AlMgZn crossover alloy used in section \ref{sec:resndis:clustering} was produced at a laboratory scale using a resistance-heated tilting furnace (Nabertherm K20/13), followed by homogenisation at 460°C for 25h and 470°C for 45h and subsequent hot and cold rolling. Full details of the production process and alloy composition are reported elsewhere by Aster \textit{et al.} \cite{aster2025clustering}. Specimens sectioned from the rolled sheet were solution heat-treated at 465°C for 35 minutes and water quenched, with the interval to any subsequent ageing treatment kept below three minutes. A pre-ageing condition (PA: 100°C/5 h) and a long-term ageing condition (LTA: 100°C/42 days) were applied.

	\subsubsection{Paint-baked AlMgZn crossover alloy}
    \noindent The PB AlMgZn crossover alloy used in section \ref{sec:resndis:PB_2pc_AlMgZn_crossalloy} was produced by induction melting, cast into a copper mould, and processed via hot and cold rolling to a final sheet thickness of 1 mm, as described in detail elsewhere (including alloy composition) by Stemper \textit{et al.} \cite{stemper2021giant}. Specimens were solution heat-treated at 550°C for 8 minutes, water quenched, and subsequently pre-aged at a temperature of 100°C for 3 h. Then, after 2\% deformation was applied, the alloy was subjected to paint-bake at 185°C for 20 min.
	
	\subsubsection{Overaged AlZnMg aerospace-grade alloy} 
    \noindent To obtain the overaged AA7075-T7 alloy reported in section \ref{sec:resndis:overaged7075}, we used a 2 mm thick AA7075-T6 sheet supplied by AMAG Austria Metall AG and further aged at 175\;°C for 24\;h. The chemical composition of the sheet (in wt.\%) was Zn 6.06, Mg 2.64, Cu 1.50, Si 0.19, Cr 0.18, Fe 0.11, with Al as balance \cite{osterreicher2020warm}.

	\subsubsection{Aerospace-grade AA2024-T3 anodised with Ce nanoparticles}
    \noindent The samples used in section \ref{sec:resndis:anodic2024} were lifted-out using a focused ion beam (FEI Versa 3D Dual Beam FIB/SEM) directly from clad AA2024-T3 alloys that were anodised in a tartaric-sulphuric acid bath (40 g$\cdot$L$^{-1}$ H$_2$SO$_4$ + 80 g$\cdot$L$^{-1}$ C$_4$H$_6$O$_6$, 37°C) at 14 V for 20 min, following established procedures already described in a previous work by Prada-Ramirez \textit{et al.} \cite{ramirez2020tartaric}. Following the anodising procedure, the samples were post-treated for 2~min in a 50~mM Ce(NO$_3$)$_3{\cdot}$6H$_2$O + 10~vol.\% H$_2$O$_2$ aqueous solution at 50°C (condition Ce10P 50C 2M in \cite{ramirez2020tartaric}) to 
	incorporate cerium-based nanoparticles into the anodic oxide layer. We emphasise that the anodic layer forms on the commercially pure aluminium cladding layer rolled onto the AA2024-T3 substrate, which results in a much more homogeneous morphology.
	
	\subsection{Magnetron-sputtering of nanocrystalline pure aluminium thin films}
	\label{sec:matnmet:rosti}
	\noindent Non-reactive magnetron-sputtering of nanocrystalline pure aluminium thin films (~100 nm thick) was performed in Ar plasma discharge at a pressure of 0.2 Pa at room temperature. A pure 3 inch diameter aluminium target with a power of 400 W, a radio-frequency substrate bias of 15 W was applied to densitfy the film. The rotational speed of the substrate holder was 1 Hz to ensure an uniform thickness. The films were prepared in a computer controlled ultra-high-vacuum deposition system model ATC 1800 (AJA International). Further details on this deposition method and system can be found elsewhere \cite{Daniel2022Acta}.
	
	\subsection{Sample preparation for electron microscopy}
	\label{sec:matnmet:sampleprep}
	
	\noindent Except for the case reported in section \ref{sec:resndis:anodic2024}, where the electron transparent samples were prepared using the conventional lift-out technique for TEM sample preparation \cite{Giannuzzi1999FIB}, the aluminium alloys used in this work were Jet Electropolished (JEP) with an electrolyte solution containing 33\% of nitric acid and 66\% of methanol (in volumetric percentage) at 12 V and -20°C. JEP was performed until perforation using a Struers Tenupol-5, starting from punched 3 mm discs with a thickness of approximately 80 \textmu m.
	
	\subsection{STEM, DPC, and EDX}
	\label{sec:matnmet:stem}
	\noindent STEM has been performed in this work using a Thermo Fisher Talos F200X equipped with an X-FEG operating at 200 keV with a nominal STEM resolution of < 0.14 nm. The STEM features a low-angle annular dark-field (LAADF) detector (DF4 designation by Thermo Fisher) which is segmented in four quadrants, allowing native acquisition of DPC micrographs as described in section \ref{sec:DPCprinciple}. The analytical EDX maps shown in sections \ref{sec:resndis:overaged7075} and \ref{sec:resndis:anodic2024} were acquired using a Super-X EDX system which features an array composed of four detectors.
	
	\subsection{Grain-size analysis with Python}
	\label{sec:matnmet:martin}

    \noindent DPC micrographs of the nanocrystalline aluminium thin films were exported as greyscale images using the log\textsubscript{10} magnitude channel in Thermo Fisher Velox, which served as input to a purpose-built grain segmentation pipeline implemented in Python. The pipeline combines a trained U-Net convolutional neural network \cite{Matthews} with SAM \cite{Kirillov2023}. Under the imaging conditions employed in this work, the U-Net alone proved sufficient for reliable grain boundary identification without recourse to the Segment Anything Model. The pipeline was originally developed for SEM micrographs \cite{DigitalSreeni} and adapted in this work for DPC input. For each detected grain, the algorithm extracts the equivalent circular diameter (ECD), aspect ratio, and circularity. Statistical outliers were removed using the criterion $Q_3 + 3 \cdot \mathrm{IQR}$, where $Q_3$ denotes the third quartile and $\mathrm{IQR}$ the interquartile range of the ECD distribution. Average grain size was determined in accordance with ASTM Standard E112 \cite{E112}, and the results are presented as grain size number, size distribution, and ECD histogram as described in section~\ref{sec:resndis:grains}.
	
	\medskip
	\noindent \textbf{Supporting Information} The image segmentation code used in section \ref{sec:matnmet:martin} was implemented from the work of Dr. Sreenivas Bhattiprolu \cite{DigitalSreeni} which is available for download on GitHub: \href{https://github.com/bnsreenu/python_for_microscopists/blob/master/373_SAM_based_ASTM_Grain_Size_Analysis.ipynb}{373\_SAM\_based\_ASTM\_Grain\_Size\_Analysis}. An youtube video explanation is also provided by Dr. Bhattiprolu: \href{https://www.youtube.com/watch?v=KCGKLh05HRM}{youtube.com/watch?v=KCGKLh05HRM}. We have also written an Python script which compiles into an executable app that allows segmenting DPC images. This script can be downloaded at: \href{https://github.com/SeSam-MUL/HSV-Wizard}{github.com/SeSam-MUL/HSV-Wizard}.

    \newpage
	\medskip 
	\noindent \textbf{Acknowledgements} 
	MAT is grateful to Dr. Lukas Stemper (AMAG Austria Metal AG) and Professor Stefan Pogatscher for past collaborations on the inception of the aluminium crossover alloys. The authors gratefully acknowledge the Austrian Research Promotion Agency (FFG) for funding the Thermo Fisher Scientific Talos F200X G2 S/TEM through the project 3DnanoAnalytics (FFG-No 858040).
	\medskip
	
	\bibliographystyle{MSP}
	\bibliography{references_doi.bib}
\end{document}